\def\br{{\bf r}}
\def\bR{{\bf R}}
\def\bk{{\bf k}}
\def\bkp{{\bk '}}
\def\bd{{\bf d}}
\def\bE{{\bf E}}
\def\w0{\omega_0}
\def\k0{k_0}
\def\wk{\omega_k}

\def\epkj{\epsilon_{\bk j}}
\def\epkjs{\epsilon_{\bk j}^\star}

\def\ekj{\hat{e}_{\bk j}}

\def\akj{a_{\bk j}}
\def\akjd{a_{\bk j}^\dagger}

\def\Fln{F_{\ell n}}
\def\Flm{F_{\ell m}}
\def\Fmp{F_{mp}}

\def\besC{\mid vac, \uparrow_C\rangle}
\def\besdC{\langle vac, \uparrow_C \mid}

\def\m{\mbox{\boldmath $\mu$}}

\documentclass[preprint,aps,showpacs]{revtex4}
\usepackage{graphicx}
\begin{document}

\title{Nonlocal field correlations and dynamical Casimir-Polder forces between
one excited- and two ground-state atoms}

\author{R. Passante, F. Persico, L. Rizzuto}
\affiliation{ CNISM and Dipartimento di Scienze Fisiche ed
Astronomiche, Universit\'{a} degli Studi di Palermo, Via Archirafi
36, I-90123 Palermo, Italy }

\email{lucia.rizzuto@fisica.unipa.it}


\begin{abstract}
The problem of nonlocality in the dynamical three-body
Casimir-Polder interaction between an initially excited and two
ground-state atoms is considered. It is shown that the nonlocal
spatial correlations of the field emitted by the excited atom
during the initial part of its spontaneous decay may become
manifest in the three-body interaction. The observability of this
new phenomenon is discussed.
\end{abstract}

\maketitle

\section{\label{sec:1}Introduction}

The existence of observable effects originating from the quantum
nature of the electromagnetic field has received much attention
since 1948 when Casimir predicted that zero-point field
fluctuations give rise to an attractive force between two neutral
conducting plates at rest in the vacuum \cite{C48}. The same year
Casimir and Polder provided an explanation for the retarded
long-range van der Waals interaction between two neutral
polarizable objects as a manifestation of the zero-point energy of
the electromagnetic field \cite{CP48}. They found that retardation
yields a decay law of the interaction energy as $r^{-7}$ at large
interatomic separation (Casimir-Polder potential). Casimir's
results also showed how geometrical constraints can affect vacuum
field fluctuations. These effects have been measured
experimentally, and the results obtained are in good agreement
with the theory \cite{SBCSH93,Lamoreaux97,BMM01}. More recently
the attention of theoreticians has been also drawn by how a
dynamical change of geometrical or topological boundaries affects
vacuum field fluctuations, giving rise to observable effects such
as modifications of the Casimir effect \cite{ILC04} or creation of
real quanta from the vacuum (the so-called dynamical Casimir
effect) \cite{FD76,KG99}. Dynamical Casimir effect is closely
related to the Unruh effect, which establishes that an atom or a
charge uniformly accelerated in the vacuum behaves as if it were
immersed in a bath of thermal radiation with a temperature
proportional to its acceleration \cite{U76}. The concept
underlying all these phenomena is that the notion of vacuum and
its physical properties depend critically on the physical system
considered and on the boundary conditions.

Initially, the interatomic Casimir-Polder (CP) potential was
understood in terms of the energy of the zero-point fluctuations
of the electromagnetic field. More recently, it was shown that it
can be also obtained as a consequence of the existence of
correlations between the fluctuating dipole moments of the atoms,
induced by the spatially correlated vacuum fluctuations. In other
words, the CP interaction energy between two atoms in their ground
state can be seen as the classical interaction energy between the
instantaneous atomic dipoles, induced and correlated by the
spatially-correlated vacuum field fluctuations \cite{PT93,PPR03a}.
This model is conceptually intriguing because gives a classical
picture of CP forces: the quantum nature of the electromagnetic
field enters only in the assumption of vacuum fluctuations as a
``real" field affecting atomic dynamics. This model has been also
generalized to the three-body CP potential between three atoms in
their ground or excited states \cite{CP97,PPR05}. In this case,
any pair of atoms interacts via their dipole moments which are
induced and correlated by the vacuum field fluctuations, modified
(dressed) by the presence of the third atom. Because the presence
of one atom modifies the spatial correlations of the electric
field, the interaction between two atoms changes if a third atom
is present, and this evantually yields a non-additive interaction.
Thus CP forces between atoms are a direct manifestation of the
existence of nonlocal correlation of zero-point fluctuations, and
so their measure can be used as an indirect evidence of field
correlations and for investigating their nonlocal properties.

Many conceptual difficulties in quantum mechanics are involved in
the notion of nonlocal correlations, also in connection with
relativistic causality. In particular, Hegerfeldt reopened the
question of nonlocality and causality in the QED context on a
quite general basis \cite{H94,MJF95,PT97,PPO00, POP01,
AKP01,CPPP03}. Specific calculations have shown that the dynamics
of local atomic or field operators is causal, but that the
correlation of atomic excitations of two spatially separated atoms
exhibits a nonlocal behaviour, being different from zero even if
the two atomic sites have a spacelike separation \cite{BCPPP90,
PT97}. Non-local terms appear also in the spatial correlations of
the energy density during the dynamical dressing/undressing of a
static source interacting with the relativistic scalar field
\cite{LP95}. The question of relativistic causality and its
relation with the nonlocal correlations of vacuum fluctuations has
been also examined in connection with the Unruh effect \cite{U84}.
All this emphasizes the conceptual importance of investigating
whether nonlocal correlations of vacuum fluctuations may be at the
origin of observable effects. Recently we have examined the
question of causality in the dynamical CP interaction between an
excited and a ground-state atom during the dynamical evolution of
the excited atom \cite{RPP04}. If one of the two atoms is in the
excited state at time $t=0$, the interaction energy between the
two atoms is non-vanishing only after the {\it causality time}
$t=R/c$, R being the interatomic distance. This indicates that
causality is a basic property of QED as far as local quantities
such as field energy densities or the two-body CP forces (which
are in principle measurable by a local observation) are
considered.

A question worth considering is what happens when nonlocal
quantities, such as the three-body forces, are considered. This is
related to the very nature of many-body forces, which appear to be
inherently nonlocal because they cannot be measured through a
single local measurement. We have recently investigated this issue
in the time-dependent three-body CP interaction between three
atoms initially in their bare ground-state, during their dynamical
self-dressing \cite{PPR06}. Our results indeed indicate that there
exist time intervals and geometrical configurations of the three
atoms for which the three-body interaction energy exhibits a
nonlocal behaviour, related to nonlocal properties of the field
correlation functions. This should allow to investigate the
nonlocal properties of the field by measurements of the
(observable) three-body Casimir-Polder forces.

In this paper we address a similar question for the dynamical CP
potential between one excited- and two ground-state atoms, during
the dynamical self-dressing of the excited atom and the initial
part of its spontaneous decay. We investigate the causality
problem in the time-dependent three-body CP potential and its
relation to the nonlocal properties of the field emitted by the
excited atom during its short-time evolution. Compared to the case
of three ground-state atoms, a resonant contribution is now
present both in the field emitted by the excited atom and in the
time-dependent potential; this new contribution arises from an
additional term in the correlation of the induced dipoles of the
two ground-state atoms.

The paper is organized as follows. In section II we consider a
two-level atom (say C) initially in its excited state and we
investigate the dynamics of the spatial correlations of the
electric field during its dynamical evolution. We show that this
correlation has a non-local behaviour. In Section III we calculate
the interaction energy between a pair of atoms located at some
distance from atom C and show that this energy has nonlocal
properties as a consequence of field nonlocality. Finally, in
section IV, we calculate the total three-body CP potential by an
appropriate symmetrization of the role of the three atoms and
discuss how the nonlocal behaviour of the field correlation
function influences the dynamical CP potential between the three
atoms.

\section{\label{sec:2}The field correlation function}

We first evaluate the correlation function of the electromagnetic
field during the spontaneous decay at short times of a two-level
atom (C), initially in its bare excited state. We describe our
system using the multipolar coupling Hamiltonian, in the Coulomb
gauge and within the dipole approximation \cite{CT84},
\begin{eqnarray}
H= H_C + H_F + H_{int} \label{eq:1}
\end{eqnarray}

\noindent where

\begin{eqnarray}
H_C &=&\hbar \w0 S_{z};  \nonumber\\
H_F&=& \sum_{\bk j} \hbar \wk \akjd\akj;
\nonumber\\
H_{int}&=& \sum_{\bk j} \biggl(\akj e^{i\bk\cdot\br_C} - \akjd
e^{-i\bk\cdot\br_C} \biggr) \nonumber \\
&\times& \biggl(\epkj S_+ - \epkjs S_- \biggr) \label{eq:2}
\end{eqnarray}
where $\w0$ is the atomic  transition frequency, $S_{z}, S_{\pm}$
are the pseudospin operators of atom C, $\akj$, $\akjd$  are the
bosonic annihilation and creation field operators and $\br_{C}$ is
the position of atom C. $\epkj$ is the coupling constant given by
\begin{equation}
\epkj = -i \left( \frac {2\pi \hbar \wk}V \right)^{1/2} \m^{C} \cdot
\ekj
\end{equation}
where $\m^{C}$ is the matrix element of the electric dipole moment
of atom C.

The initial state is assumed as the factorized state \\ $\besC$,
with the atom C in its bare excited state and the field in the
vacuum state. First we wish to evaluate on the initial state
$\besC$ the average value of the equal-time spatial correlation of
the field at two different points $\br_A$ and $\br_B$, $\langle
d_{\perp \ell} (\br_A ,t) d_{\perp m} (\br_B,t) \rangle$ (we work
in the Heisenberg representation), where
\begin{equation}
\bd_\perp (\br,t)=i\sum_{\bk j} \left( \frac {2\pi\hbar ck}{V}
\right)^{1/2} \ekj \left( \akj (t) e^{i \bk \cdot \br} - \akjd (t)
e^{i\bk\cdot\br} \right) \label{eq:3}
\end{equation}
is the transverse displacement field operator (the momentum
conjugate to the vector potential, in the multipolar coupling
scheme) which, outside the atoms, coincides with the total
(transverse plus longitudinal) electric field operator
\cite{PZ59}. From now on we shall use the symbol $\bE$ in place of
$\bd_\perp$.

Our approach closely follows that used by Power and
Thirunamachandran in \cite{PT83,PT99}. Solving Heisenberg
equations of motion for the field operators $\akj(t)$ and
$\akjd(t)$ at the second order in the coupling constant, we obtain
the following expansion for the field operator \cite{PT83,PT99}
\begin{equation}
\bE (\br,t) = \bE^{(0)}(\br,t) + \bE^{(1)}(\br,t) +
\bE^{(2)}(\br,t) \label{eq:4}
\end{equation}

The operator $\bE^{(0)}(\br,t)$ is the free-field operator at time
t, while $\bE^{(1)}(\br,t)$ and $\bE^{(2)}(\br,t)$ are
source-dependent contributions. Explicit evaluation of
(\ref{eq:4}) shows that both $\bE^{(1)}(\br,t)$ and
$\bE^{(2)}(\br,t)$, contrarily to $\bE^{(0)}(\br,t)$, contain the
Heaviside function $\theta(ct-R)$, where $R = \mid \br -\br_C
\mid$ is the distance of the observation point $\br$ from atom C,
\begin{equation}
\bE^{(1)}(\br,t) \sim \theta(ct-R) ; \hspace{20pt}
\bE^{(2)}(\br,t) \sim \theta(ct-R)
\end{equation}
This expresses a causal behaviour of the source electromagnetic
field. Hence the electric field at the second order can be
expressed as the sum of two terms: a free-field contribution,
which is independent of the presence of atom C, and a
source-dependent contribution which is strictly causal,
\begin{eqnarray}
\bE(\br,t)=\bE^{(free)}(\br,t)+\bE^{(causal)}(\br,t) \label{eq:5}
\end{eqnarray}

It should be stressed that these results have been obtained in the
multipolar coupling scheme, where the operator conjugate to the
vector potential is the transverse displacement field;  outside
the atoms, it coincides with the total electric field, which obeys
a fully retarded wave equation. In the minimal coupling scheme, on
the contrary, the conjugate momentum is the transverse electric
field, which obeys a wave equation with the transverse current
density as source term; in this scheme we would have obtained a
non-retarded solution, and electrostatic terms should be added in
order to restore a causal propagation of the field. This
illustrates the remarkable advantage of using the multipolar
coupling Hamiltonian, which is obtained from the minimal coupling
Hamiltonian by the application of the Power-Zienau transformation
\cite{PZ59}. We now evaluate the expectation value of the
correlation function of the electromagnetic field $\besdC
E_\ell(\br_A,t) E_m(\br_B,t) \besC$ at the two points $\br_A$ and
$\br_B$. Up to the second order in the electric charge $e$ and
using the expressions for the field operator given in
\cite{PT83,PT99}, this correlation function is obtained as
\begin{eqnarray}
&\ & \langle E_\ell(\br_A,t) E_m(\br_B,t)\rangle = \langle
E_\ell^{(0)}(\br_A,t)E_m^{(0)}(\br_B,t)\nonumber\\
&\ & + E_\ell^{(1)}(\br_A,t)
E_m^{(1)}(\br_B,t)+E_\ell^{(2)}(\br_A,t)E_m^{(0)}(\br_B,t)\nonumber\\
&\ &+ E_\ell^{(0)}(\br_A,t) E_m^{(2)}(\br_B,t)\rangle \label{eq:7}
\end{eqnarray}
with
\begin{widetext}
\begin{eqnarray}
&\ & \langle E_\ell^{(0)}(\br_A,t) E_m^{(0)}(\br_B,t) \rangle
=\frac{2\pi\hbar c}{V}\sum_{\bk
j}(\ekj)_{l}(\ekj)_{m}ke^{i\bk\cdot(\br_A-\br_B)};
\label{eq:7a} \\
\nonumber\\
&\ &\langle E_\ell^{(1)}(\br_A,t) E_m^{(1)}(\br_B,t) \rangle =
\mu_n^{12} \mu_p^{12}\Fln^{\beta} \frac{e^{-i\k0
\beta}}{\beta}\Fmp^{\alpha}\frac{e^{i\k0 \alpha}}{\alpha}
\theta(ct-\beta)\theta(ct-\alpha); \label{eq:7b} \\
\nonumber\\
&\ & \langle E_\ell^{(2)}(\br_A,t) E_m^{(0)}(\br_B,t) +
E_\ell^{(0)}(\br_A,t) E_m^{(2)}(\br_B,t) \rangle = \mu_n^{12}
\mu_p^{12}\Fln^{\beta} \frac{e^{i\k0
\beta}}{\beta}\Fmp^{\alpha}\frac{e^{-i\k0 \alpha}}{\alpha}
\theta(ct-\beta)\theta(ct-\alpha)\nonumber\\
&\ & -
 \frac{2\pi}{V} \Biggl\{\sum_{\bk j} k \mu_n^{21}
\mu_p^{21}(\ekj)_{l}(\ekj)_{n} e^{-i\bk \cdot \bR_{AC}}
\frac{1}{\k0+k}\Fmp^{\alpha}\frac{1}{\alpha} (e^{-ik\alpha} -
e^{i\k0\alpha}e^{-i(\k0+k)ct})\theta(ct-\beta)\nonumber\\
&\ & + \left(c.c.(A \rightleftharpoons B, \alpha \rightleftharpoons
\beta, (ln) \rightleftharpoons (m p)) \right)\Biggr\} \label{eq:7c}
\end{eqnarray}
\end{widetext}
where $\alpha=\mid \br_B - \br_C \mid$ and $\beta=\mid \br_A - \br_C
\mid$.

Eq. (\ref{eq:7a}) describes the zero-point contribution to the
field correlation function and does not play any role in the
causality problem for three-body forces we are concerned with. On
the other hand, the two terms (\ref{eq:7b}) and (\ref{eq:7c})
depend explicitly on the position of atom C. In particular, the
term (\ref{eq:7b}) arises from the retarded field emitted by atom
C at the two points $\br_A$, $\br_B$ and it is causal. This is
expected since the electric-field operator $d^{(1)}(\br,t)$ vanish
for $t<r/c$. The other contribution, which contains both the
second-order field (causal) and the free-field at time t, is
responsible for the non-local behaviour of the field correlation
function. In order to discuss this point, we partition the field
correlation function in the following form (disregarding the
free-field contribution which, as mentioned, is not relevant for
our purposes)
\begin{eqnarray}
&\ & \besdC E_\ell (\br_A,t) E_m (\br_B,t )
\besC \nonumber\\
&\ &= \langle E_\ell (\br_A, t ) E_m (\br_B, t )
\rangle_{nr}\nonumber\\
&\ & + \langle E_\ell (\br_A, t ) E_m (\br_B, t ) \rangle_{r}
\label{eq:9}
\end{eqnarray}
where
\begin{eqnarray}
&\ & \langle E_\ell(\br_A,t)E_m(\br_B,t)\rangle_{nr} \nonumber\\
&\ & = - \frac{2\pi}{V} \Biggl\{\sum_{\bk j} k \mu_n^{21}
\mu_p^{21}(\ekj)_{l}(\ekj)_{n} e^{-i\bk \cdot \bR_{AC}}\nonumber\\
&\ & \times \frac{1}{\k0+k}\Fmp^{\alpha}\frac{1}{\alpha}
(e^{-ik\alpha} -
e^{i\k0\alpha}e^{-i(\k0+k)ct})\theta(ct-\beta)\nonumber\\
&\ & + \left(c.c.(A \rightleftharpoons B, \alpha \rightleftharpoons
\beta, (ln) \rightleftharpoons (m p)) \right)\Biggr\} \label{eq:10}
\end{eqnarray}
is the {\it nonresonant} contribution to the correlation function,
and
\begin{eqnarray}
&\ & \langle E_\ell (\br_A,t) E_m (\br_B,t) \rangle_{r}
\nonumber\\
&\ & = 2\mu_n^{12} \mu_p^{12}\Fmp^{\alpha}\Fln^{\beta}
\frac{\cos\k0(\alpha -\beta)}{\alpha\beta}\theta(ct-\beta)
\nonumber\\ &\ & \times \theta(ct-\alpha)\label{eq:11}
\end{eqnarray}
is the {\it resonant} contribution, which derives from the pole at
$k=\k0$ in the frequency integration. The {\it nonresonant} term
({\ref{eq:10}) is equal but opposite in sign to that already
obtained when atom C is in the ground state \cite{PPR06}. The {\it
resonant} term is not present in the case of a ground state atom,
of course. Inspection of (\ref{eq:10}) and (\ref{eq:11}) clearly
shows that if the two points $\br_A$ and $\br_B$ are outside the
causality sphere of atom C, that is if $\alpha,\beta>ct$, the
correlation function (\ref{eq:9}) reduces to zero. When both
points $\br_A$ and $\br_B$ are inside the light-cone of atom C,
the correlation function is modified by the presence of atom C.
All this is compatible with relativistic causality, of course.
Yet, nontrivial results are obtained if just one of the two points
$\br_A$ and $\br_B$ is inside the causality sphere of atom C. For
example, when $\alpha < ct$ and $\beta > ct$ the correlation
function is modified by the presence of atom C. Moreover, this
happens whatever the distance between the two points $\br_A$ and
$\br_B$. This result indicates nonlocal features of the field
correlation function, which originate only from the non-resonant
part of the correlation function, as clearly shown by Eqs.
(\ref{eq:10}-\ref{eq:11})

\section{\label{sec:3} The three-body contribution to the dynamical Casimir-Polder interaction}

Let us now consider two more ground-state atoms, A and B, located
at points $\br_A$ and $\br_B$ respectively. We wish to evaluate
their Casimir-Polder interaction energy $\Delta E_C(A,B)$ in the
presence of atom C. Our aim is to investigate whether the nonlocal
behaviour of the field correlation function discussed in the
previous Section may reveal itself in the time-dependent
interaction energy between the two atoms. This is indeed expected
because it is known that the Casimir-Polder interaction between
two atoms depends on the vacuum field correlations evaluated at
the atomic positions \cite{PT93}. We have already discussed a
similar problem in the case of three atoms initially in their bare
ground state \cite{PPR06}; the main difference in the present case
is the presence of a resonant contribution to the correlation
function. Our approach is a generalization to the time-dependent
case of the model already used to calculate the three-body
potential with one atom excited in a time-independent approach
\cite{PPR05}. Following the same arguments used in \cite{PPR05},
to which we refer for more details, the three-body contribution to
the interaction energy between atoms A and B in the presence of
the excited atom C consists of two terms. The first is related to
the non resonant part of the correlation function and is formally
equivalent to that obtained when the three atoms initially are in
their bare ground state. The second is related to the resonant
part of the field correlation function. Thus we write the
interaction energy between A and B in the presence of C as
\begin{equation}
\Delta E_C(A,B) = \Delta E_C(A,B)^{nr} + \Delta E_C(A,B)^{r}
\label{eq:15}
\end{equation}
where the first term is a {\it non-resonant} contribution and the
second the {\it resonant} one. These two contributions are
expressed as
\begin{eqnarray}
&\ & \Delta E_C(A,B)^{nr} = \sum_{\bk j, \bk'
j'}\alpha_{A}(k)\alpha_{B}(k')\nonumber\\
&\times& \langle E_\ell (\bk j,\br_A,t)E_m(\bkp
j',\br_B,t)\rangle_{n.r.} V_{lm}(k,k',\gamma) \label{eq:12}
\end{eqnarray}
and
\begin{eqnarray}
&\ & \Delta E_C(A,B)^r = \sum_{\bk j \bkp j'} \alpha_A(k) \alpha_B(k')
\nonumber \\
&\ & \times \langle E_\ell (\bk j, \br_A,t) E_m  (\bkp j',
\br_B,t) \rangle_r V_{\ell m}^r(\k0,\gamma).\label{eq:12a}
\end{eqnarray}
$E_\ell (\bk j, \br,t)$ are the Fourier components of
$E_\ell(\br,t)$,
\begin{eqnarray}
V_{lm}(k, k',r)=-\frac 12 \Flm^\gamma \frac {1}{\gamma}(\cos
k\gamma+\cos k'\gamma) \label{eq:13}
\end{eqnarray}
is the classical potential tensor between oscillating dipoles at
frequencies $k$ and $k'$ \cite{CP97}, and
\begin{equation}
V_{\ell m}^r (\k0,\gamma) = - F_{\ell m}^{\gamma} \left( \frac
{\cos \k0 \gamma}{\gamma} \right) \label{eq:14}
\end{equation}
is the potential tensor for dipoles oscillating at the resonant
frequency $\k0$. $\gamma=\mid \br_A -\br_B \mid$ is the distance
between dipoles A and B, and $\Fln^\gamma =\left( -\nabla^2
\delta_{\ell n} + \nabla_\ell \nabla_n \right)^\gamma$ is a
differential operator acting on the variable $\gamma$. The
resonant contribution(\ref{eq:12a}) is specific to the
excited-atom case and does not appear when all atoms are in their
ground state. After lengthy algebraic calculations, we obtain the
explicit expressions of $\Delta E_C(A,B)^{nr}$ and $\Delta
E_C(A,B)^r$
\begin{widetext}
\begin{eqnarray}
&\ &\Delta E_C(A,B)^{nr}= \frac{1}{2\pi} \mu_n^{12} \mu_p^{12}
\Flm^\gamma\Fln^\beta\Fmp^\alpha\frac{1}{\alpha\beta\gamma}\int_0^\infty
dk \frac{\alpha_A(k)\alpha_B(k)}{\k0+k}
\biggl\{\left(\sin k(\beta+\gamma)+\sin k(\beta-\gamma)\right)\nonumber\\
&\ & \times e^{-ik\alpha}\theta(ct-\alpha)+\sin
k\beta\left[e^{-ik(\alpha+\gamma)} \theta(ct-(\alpha+\gamma)) +
\mbox{sgn}(\alpha-\gamma)e^{-ik\mid\alpha-\gamma\mid}\theta(ct-\mid\alpha-\gamma\mid)
\right]\nonumber\\
&\ & + c.c.\left(\alpha\rightleftharpoons\beta
\right)\biggr\}\nonumber\\
&\ & +\frac{1}{2\pi}\mu_n^{12} \mu_p^{12}
\Flm^\gamma\Fln^\beta\Fmp^\alpha\frac{1}{\alpha\beta\gamma}
\Biggl\{\alpha_B(\k0)e^{-i\w0 t} \biggl[e^{-i\k0\alpha}
\theta(ct-\alpha)\nonumber\\
&\ & \times\int_0^\infty dk \alpha_A(k)\frac{\sin
k(\beta+\gamma)+\sin k(\beta-\gamma)}{\k0+k}e^{-i\wk t}
\nonumber\\
&\ & + \left(e^{-i\k0(\alpha+\gamma)}\theta(ct-(\alpha+\gamma))
+\mbox{sgn}(\alpha-\gamma)e^{-i\k0\mid\alpha-\gamma\mid}\theta(ct-\mid\alpha-\gamma\mid)
\right) \int_0^\infty dk \alpha_A(k)\frac{\sin
k\beta}{\k0+k}e^{-i\wk
t}\biggr]\nonumber\\
&\ & + \alpha_A(\k0)e^{i\w0
t}\left(c.c.(\alpha\rightleftharpoons\beta)\right) \Biggr\}
\label{eq:16}\\
\nonumber\\
 &\ & \Delta E_C(A,B)^{r} = \mu_n^{12}
\mu_p^{12}\alpha_A(\k0)
\alpha_B(\k0)2\Re(\Fln^{\beta}\Fmp^{\alpha}\frac{e^{i\k0
(\beta-\alpha)}}{\alpha\beta}) \Flm^\gamma (-\frac{\cos
\k0\gamma}{\gamma})\theta(ct-\beta) \theta(ct-\alpha) \label{eq:17}
\end{eqnarray}
\end{widetext}

In order to investigate possible evidence of nonlocality in the
three-body Casimir-Polder interaction (\ref{eq:15}), let us
consider a few limiting cases. For $\alpha\ll ct$ and $\beta \ll
ct$ and in the limit of large times (compatibly with the
perturbative expansion we have used), the interaction energy
between A and B in their ground states reduces to the value
\begin{eqnarray}
&\ &\Delta E_C(A,B)=-\mu_n^{12} \mu_p^{12}
F_{lm}^{\gamma}F_{ln}^{\beta}F_{mp}^{\alpha}
\frac{1}{\alpha\beta\gamma}
\nonumber \\
&\ &\times \Biggl\{\, \int_{0}^{\infty} dk \frac{
\alpha_{A}(k)\alpha_{B}(k)}{k+k_0}\left[ \cos k\alpha \Big( \sin
k(\beta+\gamma)\right.\nonumber\\
&\ & \left. + \sin k(\beta-\gamma)\Big)+\sin k\beta \Big( \cos
k(\alpha+\gamma)\right.\nonumber\\
&\ & \left. + \mbox{sgn} (\alpha-\gamma)\cos k(\alpha-\gamma) \Big)
\right] + \Big( c.c.
(\alpha\rightleftharpoons \beta \Big)\Biggr\} \nonumber\\
&\ & + \mu_n^{12} \mu_p^{12}\alpha_A(\k0)
\alpha_B(\k0)\Fmp^{\alpha}\Fln^{\beta}\Flm^{\gamma}\frac{1}{\alpha\beta\gamma}
\nonumber\\
&\ &\times \left( \cos \k0 (\alpha-\beta+\gamma) + \cos \k0
(\alpha-\beta-\gamma) \right)\label{eq:18}
\end{eqnarray}
which is already known from time-independent calculations
\cite{PPR05}. This means that, after a certain time, the
interaction energy settles to a quasi-stationary value, as
expected \cite{RPP04}.

A noteworthy result is obtained when we consider a time $t$ such
that $\alpha
> ct$ and/or $\beta > ct$. This means that at least
one of the two atoms A and B is outside the causality sphere of C.
Quite unexpectedly, equations (\ref{eq:16}) and(\ref{eq:17}) show
that in this case the interaction energy between A and B is
affected by the presence of C. In order to point out the most
relevant aspects, let us focus on the specific configuration
$\alpha , \beta
> ct$ and $\gamma < ct$, that is A and B outside of the light cone
of C but inside the light cone of each other. This configuration
of the atoms and their causality spheres are schematically
illustrated in Fig. 1. The interaction energy $\Delta E_C(A,B)$ is
then
\begin{eqnarray}
&\ &\Delta E_C(A,B)= \frac{1}{2\pi} \mu_n^{12} \mu_p^{12}
F_{lm}^{\gamma}F_{ln}^{\beta}F_{mp}^{\alpha} \frac{1}{\alpha \beta
\gamma} \nonumber\\
&\ &\times \Re \left\{\int_{0}^{+\infty} \!
\alpha_{A}(k)\alpha_{B}(k) \frac{\sin k\alpha}{\k0 + k}\Biggl( \cos
k(\beta -\gamma )\right.
\nonumber \\
&\ & - \left. \frac 12 \mbox{sgn}(\beta -\gamma-ct) e^{-i k(\beta
-\gamma )} - \frac 12 e^{ik(\beta -\gamma)} \Biggr) -
\alpha_{B}(\k0) \right.
\nonumber \\
&\ & \times \left. \int_{0}^{+\infty}  \! dk \alpha_{A}(k)
\frac{\sin k\alpha}{\k0 +k} e^{-i(\wk+\w0) t}\Biggl( \cos \k0 (\beta
-\gamma)\right.
\nonumber \\
&\ & - \left. \frac 12 e^{-i\k0 (\beta -\gamma)} - \frac 12
\mbox{sgn}(\beta -\gamma-ct) e^{i\k0 (\beta -\gamma)} \Biggr)
\right.\nonumber\\
&\ & + \left. \Biggl( c.c.(A \rightleftharpoons B, \alpha
\rightleftharpoons \beta ) \Biggr) \right\} \label{eq:19}
\end{eqnarray}
The main point is that there are time intervals for which the
expression above does not vanish: this happens when $ct<\alpha
<\gamma +ct$ and/or $ct<\beta <\gamma +ct$. We stress that in such
cases both atoms A and B are outside the light come of C:
nonetheless their Casimir-Polder interaction energy is affected by
atom C, indicating nonlocal aspects in their interaction energy.
This does not contradict the fact that in this case the
correlation function (\ref{eq:9}) can be zero if both A and B are
outside the light-cone of C. In fact, the calculation of the
quantity $\Delta E_C(A,B)$ involves a sum over the field modes of
a product of the electric field Fourier components and of the
interaction potential $V_{lm}$, which also depends on $k$. We also
observe that this effect derives exclusively from the non-resonant
contributions to the three-body CP potential: the {\it resonant}
three-body CP potential is non-vanishing only when both atoms $A$
and $B$ are inside the causality sphere of C. Thus it seems that
the nonlocal properties of the electromagnetic field emitted by
atom C during its dynamical self-dressing become manifest in the
time-dependence of the three-body CP interaction energy between
atoms $A$ and $B$, and only the (nonresonant) virtual processes
contribute to this effect.

An important conceptual point is the physical meaning of the
interaction energy $\Delta E_C(A,B)$. It is not a potential energy
related to a single atom or to the whole system of the three
atoms, but it is related to the change of the interaction between
two atoms (A and B) due to the third atom (C). Therefore its
measurement must necessarily involve some {\it correlated}
measurements on both atoms A and B, in order to separate it from
other contributions to the three-body energy such as $\Delta
E_B(A,C)$ and $\Delta E_A(B,C)$.

\begin{figure}[ht]
\begin{center}
\includegraphics*[width=12cm]{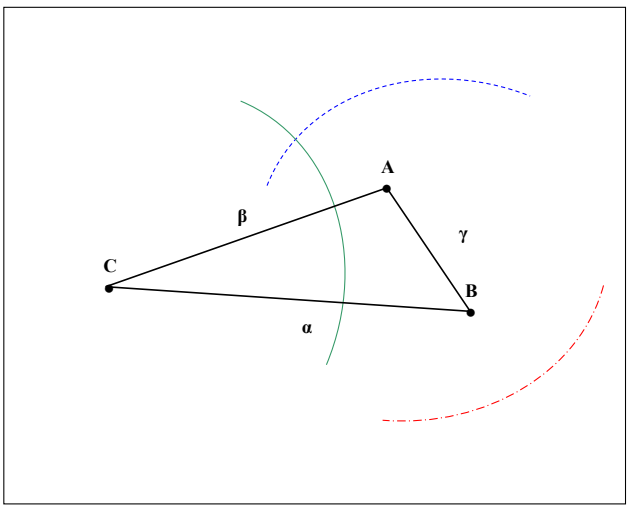}
\end{center}
\caption{A configuration of the three atoms A, B, C at time $t$
such that $\alpha > ct$, $\beta > ct$ and $\gamma <ct$, for which
a nonlocal behaviour of the interaction energy $\Delta E_C(A,B)$
is found. Red, blue and green (long-dash-dot, dashed and
continuous lines, respectively) circumference's arcs of radius
$ct$ represent respectively the causality spheres of atoms A, B
and C at time $t$.} \label{Figure1}
\end{figure}

\section{\label{sec:4} The time-dependent three-body CP potential
between atoms A,B and C}

We now evaluate the following quantity, obtained by a
symmetrization of the interaction energies of any pairs of atoms
in the presence of the third one
\begin{eqnarray}
&\ & \Delta E(A,B,C) \nonumber \\
&\ & = \frac 23 \Biggl( \sum_{\bk j \bkp j'} \alpha_A(k)
\alpha_B(k')\nonumber\\
&\ &\times \langle E_\ell(\bk j,\br_A,t) E_m  (\bkp j', \br_B,t)
\rangle_{nr} V_{\ell m}(k,k',\gamma)\nonumber\\
&\ & +(A \rightarrow B \rightarrow C) \Biggr)
\nonumber \\
&\ & + \sum_{\bk j \bkp j'} \alpha_A(k) \alpha_B(k') \langle
E_\ell (\bk j, \br_A,t) E_m  (\bkp j', \br_B,t) \rangle_r\nonumber\\
&\ & \times V_{\ell m}^r(\k0, \gamma) \label{eq:20}
\end{eqnarray}
where $(A \rightarrow B \rightarrow C)$ indicates terms obtained
from the first double sum by a permutation of the atomic indices.
In stationary cases this quantity has been shown to be equivalent
to the three-body Casimir-Polder potential, as obtained by
sixth-order perturbation theory \cite{PPR05}. This motivates our
choice to consider this physical quantity. We stress that we
symmetrize only on the nonresonant part, for which the role of the
three atoms is indeed symmetrical; the resonant part should not be
symmetrized because the contribution of the three atoms to this
term is not symmetrical, only C being in an excited state.

We now wish to investigate if the nonlocal aspect discussed above
for the interaction energy $\Delta E_C(A,B)$ are present in
$\Delta E(A,B,C)$ too. Explicit evaluation of (\ref{eq:20}),
yields
\begin{eqnarray}
&\ &\Delta E(A,B,C)= \Delta E_{(I)}(A,B,C)+\Delta
E_{(II)}(A,B,C)\nonumber\\
&\ & + \Delta E_{(r)}(A,B,C) \label{eq:21}
\end{eqnarray}
where
\begin{widetext}
\begin{eqnarray}
&\ &\Delta E_{(I)}(A,B,C)= \frac{\hbar c}{12\pi}
\Fln^\alpha\Fmp^\beta\Flm^\gamma \frac{1}{\alpha\beta\gamma} \,
\Biggl\{\, \int_0^\infty\,\, du \alpha_A(iu)
\alpha_B(iu) \alpha_C(iu)\biggl[ e^{-u(\alpha+\beta+\gamma)}\biggl(6  \nonumber\\
&\ & -\,\, \mbox{sgn}(\alpha-ct) - \mbox{sgn}(\beta-ct) -
\mbox{sgn}(\gamma-ct) - \mbox{sgn}(\alpha+\beta-ct)-
\mbox{sgn}(\alpha+\gamma-ct)\nonumber\\
&\ & - \mbox{sgn}(\gamma+\beta-ct)\biggr
)\nonumber\\
&\ &  +\,\, e^{-u(\alpha-\beta+\gamma)}\biggl (-
\mbox{sgn}(\beta-ct)+\mbox{sgn}(\alpha+\gamma-ct)\biggr) +
e^{-u(\alpha+\beta-\gamma)}\biggl(- \mbox{sgn}(\gamma-ct)
\nonumber\\
&\ & + \mbox{sgn}(\alpha+\beta-ct)\biggr)\nonumber\\
&\ &  +\,\, e^{-u(-\alpha+\beta+\gamma)}\biggl(-
\mbox{sgn}(\alpha-ct) + \mbox{sgn}(\beta+\gamma-ct)\biggr) \biggr]
\Biggr\}; \label{eq:22}
\end{eqnarray}
and
\begin{eqnarray}
&\ &\Delta E_{(II)}(A,B,C)= \frac{\hbar c}{12\pi}
\Fln^\beta\Fmp^\alpha\Flm^\gamma\frac{1}{\alpha\beta\gamma} \,
\Biggl\{2\alpha_B(\k0)\biggl[ \cos\k0(\alpha-ct)\int_0^\infty\,du
\,\alpha_A(iu) \alpha_C(iu)
\nonumber\\
&\ & \times \biggl(e^{-u(\beta+\gamma+ct)} +
\mbox{sgn}(\beta+\gamma-ct)
e^{-u\vert\beta+\gamma-ct\vert}+\mbox{sgn}(\beta-\gamma-ct)
e^{-u\vert\beta-\gamma-ct\vert}\nonumber\\
&\ & +\mbox{sgn}(\beta-\gamma+ct)
e^{-u\vert\beta-\gamma+ct\vert}\biggr)\nonumber\\
&\ & +\,\, \frac {\sin\k0(\alpha-ct)}{\k0} \int_0^\infty\,du\,\,
u\,\alpha_A(iu) \alpha_C(iu)\biggl(e^{-u(\beta+\gamma+ct)}+
e^{-u\vert\beta+\gamma-ct\vert} + e^{-u\vert\beta-\gamma-ct\vert}-
e^{-u\vert\beta-\gamma+ct\vert}\biggr)\nonumber\\
&\ & +\, (B\rightleftharpoons C,\, \beta\rightleftharpoons\gamma
)\biggr]\theta(ct-\alpha) + 4\alpha_A(\k0)
\biggl[\cos\k0(\beta+\gamma-ct)\int_0^\infty
du\,\alpha_B(iu)\alpha_C(iu) \Big( e^{-u(\alpha+ct)}
\nonumber\\
&\ & +\,\, \mbox{sgn}(\alpha-ct)e^{-u\vert \alpha-ct\vert} \Big)+
\frac{\sin\k0(\beta+\gamma-ct)}{\k0} \int_0^\infty
du\,u\,\alpha_B(iu)\alpha_C(iu)\Big( e^{-u(\alpha+ct)}
+ e^{-u\vert \alpha-ct\vert}\Big)\biggr]\nonumber\\
&\ & \times\theta(ct-(\beta+\gamma))\nonumber\\
&\ & +\,\, (A\rightleftharpoons
B,\,\,\alpha\rightleftharpoons\beta )+ (A\rightleftharpoons
C,\,\,\alpha\rightleftharpoons\gamma )\Biggr\}; \label{eq:23}
\end{eqnarray}
are the nonresonant contributions, while
\begin{equation}
\Delta E(A,B,C)_{(r)}=-\mu_n^{12} \mu_p^{12}\alpha_A(\k0)
\alpha_B(\k0)2\Re(\Fln^{\beta}\Fmp^{\alpha}\frac{e^{i\k0
(\beta-\alpha)}}{\alpha\beta})\Flm^\gamma (\frac{\cos
\k0\gamma}{\gamma}) \theta(ct-\beta) \theta(ct-\alpha)
\label{eq:24}
\end{equation}
\end{widetext}
is the resonant one. We have assumed isotropic atoms, that is
$\m_i\m_j=\frac 1 3\vert \m\vert^2$, and  $\alpha(iu)$ is the
dynamic polarizability extended to imaginary frequencies.

Eq. (\ref{eq:21}) describes the time-dependent symmetrized
three-body CP potential as a function of time for a generic
configuration of the three atoms (for times shorter than the
spontaneous decay time of the excited atom, due to the limitations
of our perturbative treatment). As in the case discussed in the
previous Section, we now consider specific cases relevant for our
discussion about nonlocal aspects of the dynamical interaction
energy. First of all, it is immediate to see that $\Delta
E(A,B,C)$ vanishes if each atom is outside the light cone of the
other two, that is for $\alpha,\beta,\gamma > ct$. On the
contrary, $\Delta E(A,B,C)$ is non-vanishing for times such that
each atom is separated by a time-like interval from the other two.
In particular, for large times, the time-dependent terms rapidly
decrease to zero and we find the well-known stationary result
\cite{PPR05}. This means that after a transient characterized by a
time-dependent interaction, the three-body interaction energy
settles to the time-independent Casimir-Polder interaction between
three atoms with an excited atoms. All these results are
compatible with relativistic causality and similar to those
previously obtained for the dynamical three-body Casimir-Polder
interaction between three ground-state atoms \cite{PPR06}.
However, when the spatial configuration of the three atoms is such
that two of them are separated by a time-like distance we find a
non-vanishing three-body interaction, even if the third atom is
outside their causality sphere. For example, when the separation
of A from the other two atoms is space-like, Eq.(\ref{eq:21})
yields
\begin{widetext}
\begin{eqnarray}
&\ & \Delta E(A,B,C) = \frac{\hbar c}{6\pi}
\Fmp^\alpha\Fln^\beta\Flm^\gamma \frac{1}{\alpha\beta\gamma}
\int_{0}^{\infty} \! du\!
\alpha_{A}(iu)\alpha_{B}(iu)\alpha_{C}(iu)(e^{-u(\alpha+\beta+\gamma)}+e^{-u(-\alpha+\beta+\gamma)})
\nonumber\\
&\ & -\frac{1}{6\pi} \mu^{12}_{n}
\mu^{12}_{p}\Fmp^\alpha\Fln^\beta\Flm^\gamma
\frac{1}{\alpha\beta\gamma}2\Re \Biggl\{\, \alpha_{B}(\k0)
e^{-i\k0(\alpha -ct)}\int_{0}^{\infty} \,dk\, \alpha_{A}(k)
\frac{\sin
k(\beta+\gamma)}{\k0+k}e^{i\wk t}\Biggr\}\nonumber\\
&\ & -\frac{1}{6\pi} \mu_n^{12} \mu_p^{12}
\Fmp^\alpha\Fln^\beta\Flm^\gamma \frac{1}{\alpha\beta\gamma} 2\Re
\Biggl\{\, \alpha_{C}(\k0)e^{-i\k0(\alpha-ct)} \int_{0}^{\infty}
\,dk\, \alpha_{A}(k) \frac{\sin k(\beta+\gamma)}{\k0+k}e^{i\wk
t}\Biggr\} \label{eq:25}
\end{eqnarray}
\end{widetext}
\noindent which in general does not vanish. Thus the nonlocal
features of field emitted by the atoms during their self-dressing
are evident also in the three-body interaction energy $\Delta
E(A,B,C)$, with features which may differ from to those of $\Delta
E_C(A,B)$. Eq.(\ref{eq:25}) shows also that the nonlocal features
of $\Delta E(A,B,C)$ stem from the non-resonant contributions, so
that they are exclusively due to the virtual photons dressing the
atoms, as we have recently discussed in the case of ground-state
atoms \cite{PPR06}.

\section{\label{sec:5} Conclusions}
We have considered the Casimir-Polder interaction energy between
three atoms with one atom initially in its excited state, using a
time-dependent approach. We have discussed the problem of
relativistic causality in the interaction between the atoms and
its connection with the non-locality of spatial field
correlations. The spatial correlation function of the field
emitted during the spontaneous decay of the excited atom has been
first obtained. We have shown that a non-local behaviour appears,
in agreement with previous results, and that it is related to a
non-resonant contribution related to the emission of virtual
photons. We have shown that a non-local behaviour appears also in
the dynamical Casimir-Polder interaction between two other
ground-state atoms, during the initial stage of the spontaneous
decay of the first atom. We have suggested that the appearance of
this non-local behaviour can be ascribed to the non-locality of
the field correlation function and that this new phenomenon should
be observable. Thus we conclude that the nonlocal properties of
the electromagnetic field emitted by the atoms during their
dynamical self-dressing may become manifest in the time-dependence
of the Casimir-Polder potential. We remark that previous studies
of causality in the time-dependent two-body Casimir-Polder
interaction have not shown indications of non local behaviour
\cite{RPP04}. Hence the causality problem appears quite more
complicated and subtle in the case of the time-dependent
three-body Casimir-Polder energy, where nonlocal aspects may
become manifest.

\begin{acknowledgments}
The authors wish to thank T. Thirunamachandran for valuable
discussions about the subject of this paper. This work was in part
supported by the bilateral Italian-Belgian project on
``Casimir-Polder forces, Casimir effect and their fluctuations"
and by the bilateral Italian-Japanese project 15C1 on ``Quantum
Information and Computation" of the Italian Ministry for Foreign
Affairs. Partial support by Ministero dell'Universit\`{a} e della
Ricerca Scientifica e Tecnologica and by Comitato Regionale di
Ricerche Nucleari e di Struttura della Materia is also
acknowledged.
\end{acknowledgments}


\begin{references}
\bibitem{C48} H.B.G. Casimir, Proc. K. Ned. Akad. Wet. B {\bf 51}
(1948) 793
\bibitem{CP48} H.B.G. Casimir, D. Polder, Phys. Rev. {\bf 73} (1948)
360
\bibitem{SBCSH93} C.I. Sukenik, M.G. Boshier, D. Cho, V. Sandoghdar, E.A. Hinds,
Phys.\ Rev.\ Lett.\ {\bf 70} (1993) 560.
\bibitem{Lamoreaux97} S.K. Lamoreaux, Phys. Rev. Lett. {\bf 78}, 5
(1997)
\bibitem{BMM01} M. Bordag, U. Mohideen, V.M. Mostepanenko, Phys.\
Rep.\ {\bf 353}, 1 (2001).
\bibitem{ILC04} D. Iannuzzi, M. Lisanti, F. Capasso, Proc. Nat. Ac. Sci. {\bf
101}, 4019 (2004)
\bibitem{FD76}S.A. Fulling and P.C.W. Davies, Proc.\ R.\ Soc.\ London\ A
{\bf 348}, 393 (1976).
\bibitem{KG99} M. Kardar and R. Golestanian, Rev.\ Mod.\ Phys.\ {\bf 71}, 1233 (1999).
\bibitem{U76} W.G. Unruh, Phys.\ Rev.\ D\ {\bf 14}, 870 (1976).
\bibitem{PT93} E.A. Power, T. Thirunamachandran, Phys.\ Rev.\ A\ {\bf 48} (1993)
4761.
\bibitem{PPR03a} R. Passante, F. Persico, L. Rizzuto, Phys. Lett. A {\bf 316}, 29 (2003)
\bibitem{CP97} M. Cirone, R. Passante,
J. Phys. B: At. Mol. Opt. Phys. {\bf 40}, 5579 (1997)
\bibitem{PPR05} R. Passante, F. Persico, and L. Rizzuto, J. Mod. Opt. {\bf 52}, 1957 (2005)
\bibitem{H94} C.G. Hegerfeldt, Phys. Rev. Lett. {\bf 72}, 596 (1994)
\bibitem{CPPP03} G. Compagno, G. M. Palma, R. Passante, and F.
Persico, in {\it The Physics of Communication: Proceeding of the
XXII nd Solvay Conference in Physics}, I. Antoniou, V.A. Sadovnichy,
H. Walter eds., World Scientific, Singapore, 2003, p. 389
\bibitem{MJF95} P. W. Milonni, D. F. V. James, and H. Fearn,
Phys. Rev. A. \textbf{52}, 1525 (1995)
\bibitem{PT97} E. A. Power, T. Thirunamachandran, Phys. Rev. A \textbf{56}, 3395 (1997)
\bibitem{PPO00} T. Petrosky, G. Ordonez, I. Prigogine,
Phys. Rev. A \textbf{62}, 42106 (2000)
\bibitem{POP01} T. Petrosky, G. Ordonez, I. Prigogine,
Phys. Rev. A \textbf{64}, 62101 (2001)
\bibitem{AKP01} I. Antoniou, E. Karpov, G. Pronko,
Found. of Phys. \textbf{31}, 1641 (2001)
\bibitem{BCPPP90} A. K. Biswas, G. Compagno, G. M. Palma, R. Passante, and F. Persico,
Phys. Rev. A. \textbf{42}, 4291 (1990)
\bibitem{LP95} A. La Barbera, R. Passante, Phys. Lett. A {\bf 206}, 1 (1995)
\bibitem{U84} W. G. Unruh, R. M. Wald, Phys.\ Rev.\ D\ {\bf 29},
1047 (1984).
\bibitem{RPP04} L. Rizzuto, R. Passante, F. Persico, Phys. Rev. A {\bf 70}, 012107 (2004)
\bibitem{PPR06} R. Passante, F. Persico, and L. Rizzuto, J. Phys. B. {\bf 39}, S685 (2006)
\bibitem{CT84} D. P. Craig and T. Thirunamachandran,
{\it Molecular Quantum electrodynamics}, Academic Press
(1984)
\bibitem{PZ59} E.A. Power, S. Zienau, Phil. Trans. Roy. Soc.
London A {\bf 251}, 427 (1959)
\bibitem{PT83} E. A. Power, T. Thirunamachandran, Phys. Rev. A {\bf 28}, 2663 (1983)
\bibitem{PT99} E. A. Power, T. Thirunamachandran, Phys. Rev. A {\bf 60}, 4927 (1999)
\end{references}
\end{document}